\documentclass[a4paper]{article}
\usepackage{Odyssey2018}
\usepackage{amsopn}
\usepackage{epsfig,amssymb,amsmath}
\usepackage{multirow}
\ninept

\title{Deep Discriminant Analysis for i-vector Based Robust Speaker Recognition}

\name{Shuai Wang, Zili Huang, Yanmin Qian, Kai Yu}

\address{    Key Laboratory of Shanghai Education Commission for Intelligent Interaction and Cognitive Engineering \\
    SpeechLab, Department of Computer Science and Engineering \\
    Brain Science and  Technology Research Center \\
    Shanghai Jiao Tong University, Shanghai, China \\
    \thanks{The corresponding author is Kai Yu} \thanks{This work has been supported by the National Key Research and Development Program of China under Grant No.2017YFB1002102 and the China NSFC projects (No. U1736202 and No. 61603252). Experiments have been carried out on the PI supercomputer at Shanghai Jiao Tong University}
\small \tt \{feixiang121976,huangziliandy, yanminqian, kai.yu\}@sjtu.edu.cn}

\begin{document}
\maketitle

\begin{abstract}
Linear Discriminant Analysis (LDA) has been used as a standard post-processing procedure in many state-of-the-art speaker recognition tasks. Through maximizing the inter-speaker difference and minimizing the intra-speaker variation, LDA projects {\it i}-vectors to a lower-dimensional and more discriminative subspace. In this paper, we propose a neural network based compensation scheme(termed as deep discriminant analysis, DDA) for {\it i}-vector based speaker recognition, which shares the spirit with LDA. Optimized against softmax loss and center loss at the same time, the proposed method learns a more compact and discriminative embedding space. Compared with the Gaussian distribution assumption of data and the learnt linear projection in LDA, the proposed method doesn't pose any assumptions on data and can learn a non-linear projection function. Experiments are carried out on a short-duration text-independent dataset based on the SRE Corpus, noticeable performance improvement can be observed against the normal LDA or PLDA methods.
\end{abstract}
\section{Introduction}
Speaker Recognition aims to recognize or verify a speaker's identity through the given speech segment. Since proposed in \cite{dehak2011front}, {\it i}-vector has become the state-of-the-art speaker modeling technique, it is a simple but elegant factor analysis model, inspired by the Joint Factor Analysis (JFA) \cite{kenny2005joint} framework. Though some researchers have been working on improving the {\it i}-vector model itself\cite{lei2014novel,rao2015normalization}, more researchers pay attention to the compensation techniques  in the {\it i}-vector space\cite{garcia2011analysis,sadjadi2014nearest,mahto2017vector,guo2017cnn}. JFA can be regarded as a compensation method in the GMM super-vector space, which models the speaker and channel variabilities separately. {\it i}-vector simplifies JFA by modeling the speaker- and channel-dependent factors in a single low dimensional space, leaving the compensation mechanisms to the following steps. In real applications, nuisance attributes such as channel, noise can pose a huge impact on the system performance, compensation methods become necessary and have attracted more and more interest. 

Linear Discriminant Analysis (LDA) \cite{balakrishnama1998linear,nasrabadi2007pattern} is widely used in pattern recognition tasks\cite{belhumeur1997eigenfaces,haeb1992linear} to project features onto a lower-dimensional and more discriminative space. The transformation is learned via maximizing the between-class (inter-speaker) difference and minimizing the within-class (intra-speaker) variation. LDA is a simple linear transformation which is used as a preprocessor to generate reduced dimensional and channel compensated embeddings from the original {\it i}-vectors in many speaker verification systems, results on standard datasets such as the Speaker Recognition Evaluation (SRE) corpus show its effectiveness\cite{sadjadi2016ibm}. Despite its effectiveness and popularity, LDA has its limitations. For instance, LDA can provide at most $C-1$ discriminant features, where $C$ is the number of classes. It's a linear projection which may not be capable of dealing with highly non-linear separable data. Several methods such as weighted LDA \cite{kanagasundaram2012weighted} and nonparametric discriminant analysis (NDA) \cite{fukunaga1983nonparametric,sadjadi2014nearest} are proposed as a substitution of LDA in speaker verification tasks. NDA redefines the between-class scatter matrix, the expected values that represent the global information about each class are replaced with local sample averages computed based on the $k$-NN of individual samples.
Another most popular compensation method in the {\it i}-vector space is Probabilistic Linear Discriminant Analysis (PLDA) \cite{prince2007probabilistic}. It's usually used as a scoring method and combined with other compensation methods, such as LDA. Considering different scenarios, PLDA has several variations such as two-covariance PLDA\cite{brummer2010speaker}, simplified PLDA\cite{garcia2011analysis,kenny2010bayesian} and Heavy-Tailed PLDA\cite{kenny2010bayesian}. Currently, the {\it i}-vector/PLDA system achieves the state-of-the-art performance.
 
Recently, there are also some attempts using Deep Learning (DL) techniques for de-noising and channel compensation in speaker recognition. This compensation can be performed in the cepstral feature space or the {\it i}-vector space. Authors in \cite{richardson2016channel} used
features estimated by the denoising DNN as the input to an {\it i}-vector system for channel robust speaker recognition.
Authors in \cite{mahto2017vector} proposed to use an auto-encoder to learn a projection which maps noisy {\it i}-vectors to de-noised ones. To address the short-duration problem of {\it i}-vector\cite{kanagasundaram2011vector}, a Convolutional Neural Network (CNN) based system was trained in \cite{guo2017cnn} to map the {\it i}-vectors extracted from short utterances to the corresponding long-utterance {\it i}-vectors.

In this paper, we propose a discriminative neural network (NN) based compensation method in the {\it i}-vector space. The proposed NN-based method shares the same spirit with LDA, it is trained to minimize softmax loss and center loss \cite{wen2016discriminative} simultaneously, where the former forces the transformed embeddings from different classes staying apart and the latter pulls the embeddings from the same class close to their centers. With the joint supervision of softmax loss and center loss, the NN produces a projection function similar to LDA, enlarging the between-class difference and reducing the within-class variation. The proposed NN-based compensation method will be referred to as Deep Discriminant Analysis (DDA) in this paper.

The rest of the paper is organized as follows. Section 2 briefly introduces the {\it i}-vector framework. Section 3 reviews two conventional compensation methods in {\it i}-vector space, LDA and PLDA. We propose the discriminative neural network based compensation method (DDA) in Section 4, followed by experiments and results analysis in Section 5. Section 6 concludes this paper.

\section{{\it i}-vector}
Joint Factor Analysis (JFA) framework\cite{kenny2005joint} was proposed as a compensation method in the GMM super-vector space, it models speaker and channel factors in separate subspaces. The following {\it i}-vector simplifies the JFA framework by modeling a single total variability subspace\cite{dehak2011front}.
In the {\it i}-vector framework, the speaker- and session-dependent super-vector $\mathbf{M}$ (derived from UBM) is modeled as 
\begin{equation}
\label{equ:ivec}
\mathbf{M}=\mathbf{m}+\mathbf{Tx}+\mathbf{\epsilon}
\end{equation}
where $\mathbf{m}$ is a speaker and session-independent super-vector, $\mathbf{T}$ is a low rank matrix which captures speaker and session variability, $\mathbf{x} \sim \mathcal{N}(\mathbf{0}, \mathbf{I})$, is a multivariate random variable, and the termed {\it i}-vector is the posterior mean of $\mathbf{x}$. $\mathbf{\epsilon}\sim \mathcal{N}(\mathbf{0}, \mathbf{I})$, is the residual noise term to account for the variability not captured by $\mathbf{T}$.

As shown in Equation\ref{equ:ivec}, {\it i}-vector is a simple and elegant representation, which follows the standard Factor Analysis (FA) scheme. However, since {\it i}-vector contains the speaker- and channel-dependent factors in the same subspace, further channel compensation methods such as LDA are often applied to annihilate the impact of nuisance attributes.
\section{Conventional Compensation Methods}
Speaker recognition systems are fragile to noise, channel and many other factors. Compensation technologies have been heavily researched on during the past several decades. In this section, we mainly discuss the compensation methods in the {\it i}-vector space. Two methods, LDA and PLDA will be specifically introduced.

\subsection{Linear Discriminative Analysis (LDA)}
LDA is widely used in pattern recognition tasks such as image recognition\cite{belhumeur1997eigenfaces} and speaker recognition\cite{sadjadi2016ibm}. LDA calculates a matrix $\mathbf{W}$ that projects high dimensional feature vectors $\mathbf{x}$ ({\it i}-vectors in this paper) into a lower-dimensional and more discriminative subspace ($\mathbf{W}:\mathbb{R}^h \mapsto \mathbb{R}^l $). The projection can be represented as:

\begin{equation}\label{eq:ldaproj}
\mathbf{y}=\mathbf{W}^T\mathbf{x}
\end{equation}
where $\mathbf{y}$ denotes the compensated embedding and $\mathbf{W}$ is a rectangular matrix of shape $h \times l$. $\mathbf{W}$ is determined by 

\begin{align}
\hat{\mathbf{W}}&=\underset{\mathbf{W}}{\arg\max}\,\frac{\text{tr}(\mathbf{W}^T\mathbf{S}_b\mathbf{W})} {\text{tr}(\mathbf{W}^T\mathbf{S_wW})} \\
&=\underset{\mathbf{W}^T\mathbf{S}_w\mathbf{W}=\mathbf{I}}{\arg\max}\,[\text{tr}(\mathbf{W}^T\mathbf{S}_b\mathbf{W})]
\end{align}

The between-class and within-class covariance matrices, $\mathbf{S}_b$ and $\mathbf{S}_w$ respectively, can be computed as

\begin{align}
\mathbf{S_b} &= \frac{1}{N}\sum_{s=1}^SN_s(\boldsymbol{\mu}_s-\boldsymbol{\mu})(\boldsymbol{\mu}_s-\boldsymbol{\mu})^T  \\
\mathbf{S_w} &= \frac{1}{N}\sum_{s=1}^S\sum_{i=1}^{N_s}(\mathbf{x}_s^i-\boldsymbol{\mu}_s)(\mathbf{x}_s^i-\boldsymbol{\mu}_s)^T 
\end{align}
where $S$ represents the total number of speakers, $N$ represents the total number of {\it i}-vectors from all speakers. $\boldsymbol{\mu}$ represents the global mean of all $N$ {\it i}-vectors, whereas $\boldsymbol{\mu}_s$ represents the mean of {\it i}-vectors from the specific $s$-th speaker. $\mathbf{x}_s^i$ represents the $i$-th {\it i}-vector from the $s$-th speaker, $N_s$ is the number of utterances from the $s$-th speaker.

LDA has an analytical solution and the optimized $\hat{\mathbf{W}}$ is a matrix whose columns are the $l$ eigenvectors corresponding to the largest eigenvalues of $\mathbf{S}_w^{-1}\mathbf{S}_b$. However, despite its simpleness and effectiveness, LDA has several limitations, 
\leftmargini=1mm
\begin{itemize}
\item The within- and between-class matrices are formed based on Gaussian assumptions for samples of each class. If the Gaussian assumption doesn't hold, LDA is not able to learn a effective enough projection function for classification problems.
\item LDA suffers from the ``small sample size'' problem, which leads to the singularity of the within-class scatter matrix $\mathbf{S}_w$. This problem happens when the number of the samples is much smaller than the dimension of the original samples. 
\item Given the class number $C$, LDA can provide at most $C-1$ discriminant features, since the between-class scatter matrix $\mathbf{S}_b$ has of rank of $C-1$. However, this may not be sufficient for tasks in which the class number is much smaller than the dimension of input features.
\item LDA learns a linear projection function, which may not be enough for data that are highly linearly inseparable.
\end{itemize}

To address these limitations of LDA, several approaches are proposed. For instance, a nonparametric was proposed in \cite{fukunaga1983nonparametric} and applied to robust speaker verification \cite{sadjadi2014nearest,sadjadi2016ibm}. In NDA, instead of only considering the class center when  computing the between-class scatter matrix, the global information about a class is defined with local sample averages computed based on the $k$-NN of individual samples. Researchers also proposed several approaches to tackle the  ``small sample size'' problem \cite{huang2002solving,sharma2015linear}.

%
%

\subsection{Probabilistic Linear Discriminant Analysis}
{\it i}-vectors with Probabilistic Linear Discriminant Analysis (PLDA) back-end obtains the state-of-the-art performance in speaker verification. Several variants of PLDA have been introduced into the speaker verification task, including the standard PLDA\cite{prince2007probabilistic}, two-covariance PLDA\cite{brummer2010speaker}, heavy-tailed PLDA\cite{kenny2010bayesian} and the Simplified PLDA\cite{garcia2011analysis,kenny2010bayesian}. The optimization goal of all variants is to maximize the between-class difference and minimize the within-class variation. PLDA models regard {\it i}-vectors as observations from a probabilistic  generative model and can be seen as a factor analysis in the {\it i}-vector space. In our experimental settings, the variant implemented in Kaldi\cite{povey2011kaldi} achieves best performance, which we termed as Kaldi-PLDA here, it's following the formulations in \cite{ioffe2006probabilistic} and similar to the two-covariance model. 

In the Kaldi-PLDA, an {\it i}-vector $\mathbf{x}$ is assumed to be generated as,

\begin{align}
\mathbf{x} &= \boldsymbol{\mu}+ \mathbf{Au} \\
\mathbf{u} &\sim \mathcal{N}(\mathbf{v,I}) \\
\mathbf{v} &\sim \mathcal{N}(\mathbf{0,\Psi})
\end{align}

where $\mathbf{v}$ represents the class (speaker), and $\mathbf{u}$ represents a sample of that class in the projected space. Kaldi-PLDA is trained using EM algorithm, training and inference details can be found in \cite{ioffe2006probabilistic} or the Kaldi project\cite{povey2011kaldi}. In the following sections, Kaldi-PLDA will be simply referred to as PLDA.

\section{Neural Network based Approach}
\subsection{Center Loss}
Neural networks have been investigated a lot in areas such as image recognition, speech recognition\cite{hinton2012deep,dahl2012context,seide2011conversational} and speaker recognition\cite{lei2014novel,heigold2016end,snyder2016deep}. One of the most popular method is to treat the neural network as a feature extractor, whereas the learned features are called ``bottle-neck feature'' or ``deep feature''\cite{wu2015deep,yu2011improved,dvec:variani2014deep,lab:liu2015deep,chen2015multi}. For instance, in speaker recognition, researchers proposed to extract feature vectors from the last hidden layer of a well-trained speaker-discriminative DNN. In most work, the DNN is optimized against the softmax loss, which emphasizes on discriminating different speakers.

The softmax loss function is defined as 
\begin{equation}
\mathcal{L}_S = - \sum_{i=1}^{N} \log \frac{e^{\mathbf{W}^T_{s_i}\mathbf{x}_i+\mathbf{b}_{s_i}}}{\sum_{j=1}^S e^{\mathbf{W}^T_{j}\mathbf{x}_i+\mathbf{b}_{j}}}
\end{equation}
where $N$ is the total number of training samples ({\it i}-vectors), $\mathbf{x}_i$ denotes the $i$-th sample, belonging to the $s_i$-th class. $S$ is the number of softmax outputs, representing $S$ different classes. $\mathbf{W}$ is the projection weight matrix and $\mathbf{b}$ is the corresponding bias term.

Center loss \cite{wen2016discriminative} is formulated as 

\begin{equation}
\mathcal{L}_C=\frac{1}{2}\sum_{i=1}^{N}|| \mathbf{x}_i - \mathbf{c}_{s_i}||^2
\end{equation}
where $\mathbf{c}_{s_i}$ represents the center of $s_i$-th class (which the $i$-th sample belongs to) and is updated along with the training procedure.
The neural network will be trained under the joint supervision of softmax loss and center loss, formulated as,
\begin{equation}
\mathcal{L} = \mathcal{L}_S + \lambda \mathcal{L}_C
\label{eq:loss}
\end{equation}
where $\lambda$ is adopted for balancing the two loss functions. Intuitively, the softmax loss forces the learned embeddings of different classes staying apart, while the center loss pulls the embeddings from the same class close to their centers. With the joint supervision of softmax loss and center loss, the neural network learns a projection function similar to LDA, enlarging the inter-class differences and reducing the intra-class variations.

\begin{figure}[!ht]
	\centering
	\includegraphics[width=0.9\linewidth]{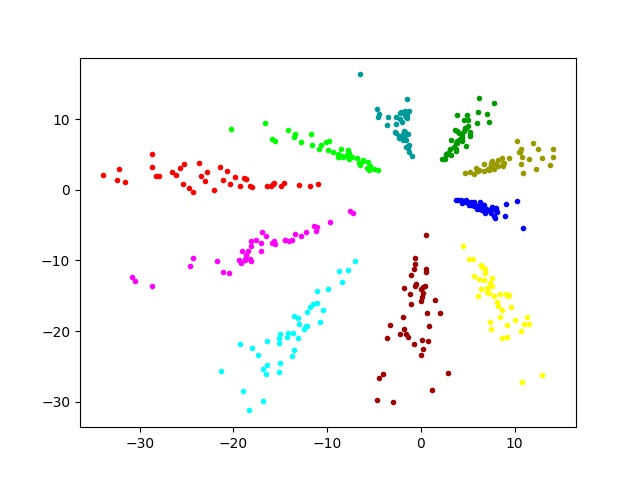}
	\caption{Embeddings supervised by softmax loss}
    \label{fig:toyce}
\end{figure}

\begin{figure}[!ht]
	\centering
	\includegraphics[width=0.9\linewidth]{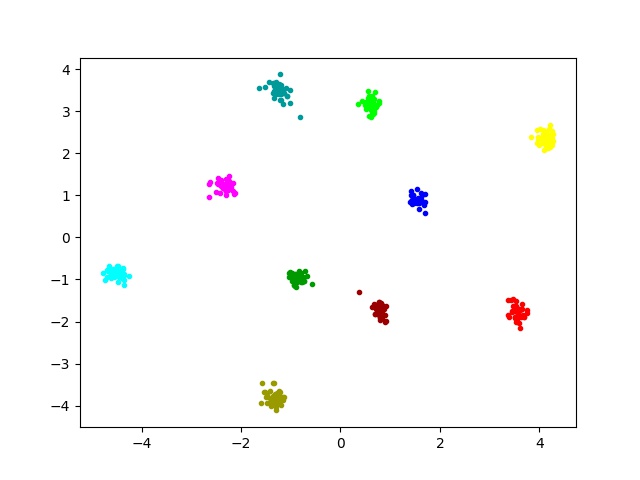}
	\caption{Embeddings supervised by softmax + center loss}
    \label{fig:toycenter}
\end{figure}

To show the effectiveness of center loss, following the approach in \cite{wen2016discriminative}, we also train a toy example on a small speaker audio dataset, which contains 10 different speakers. A 2-layer neural network is trained and the dimension of embedding layer is set as 2 for illustration. As shown in Fig.\ref{fig:toyce} and Fig.\ref{fig:toycenter} (\textbf{Best viewed in color}), with the assistance of center loss, the within-class variation reduced a lot. In the following experiments, it can be seen that this property can be generalized to scenarios where validation speakers have no overlap with the training speakers, which is the common condition in speaker verification.

\subsection{Deep Discriminant Analysis}
\begin{figure}[!ht]
	\centering
	\includegraphics[width=0.9\linewidth]{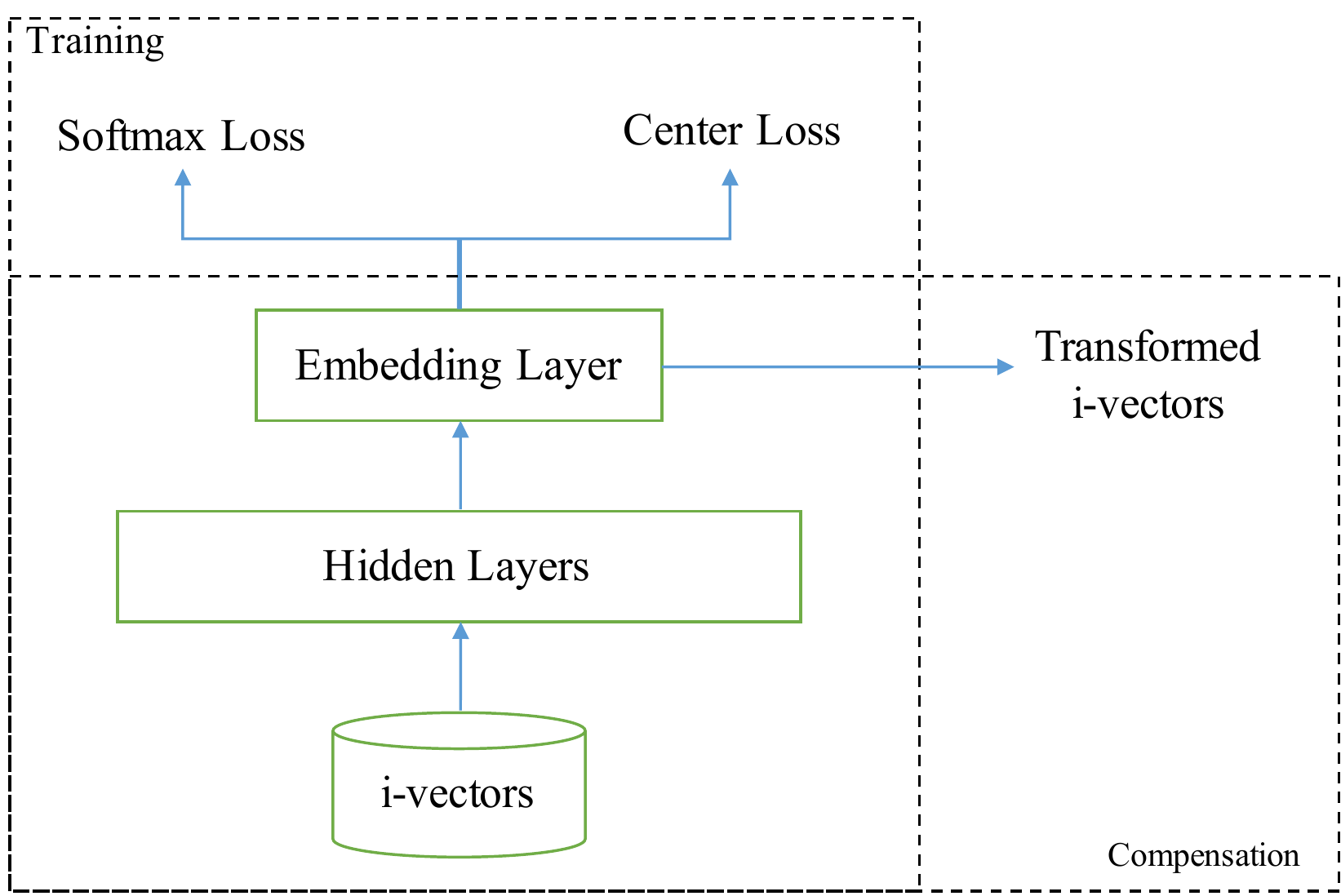}
	\caption{Architecture of DDA}
    \label{fig:system}
\end{figure}

Deep Neural Network (DNN) shows its extraordinary capability in speech recognition and speaker recognition, there is no prior assumption on the input data. Through substituting Gaussian Mixture Model (GMM) to DNN\cite{hinton2012deep}, the DNN-HMM systems achieve noticeable performance improvement compared to traditional GMM-HMM systems, which also holds for speaker recognition tasks when updating GMM-{\it i}-vector to DNN-{\it i}-vector\cite{lei2014novel}. In this section, a DNN is used to perform the channel compensation in the {\it i}-vector space. The whole architecture is depicted in Fig.\ref{fig:system}. In the training phase, the extracted {\it i}-vector from different speakers are prepared as input, the DNN is joint supervised by the softmax loss and the center loss. The last layer before the loss layer is an embedding layer, from which we extract the transformed embeddings. In the compensation stage, the source {\it i}-vectors are mapped to their corresponding transformed version through the trained neural network. Similar to the projection in Equation \ref{eq:ldaproj}, given the original {\it i}-vector $\mathbf{x}$, the compensated lower-dimensional embedding $\mathbf{y}$ can be represented as
\begin{equation}
\mathbf{y} = \mathcal{G}(\mathbf{x})
\end{equation}
where $\mathcal{G}()$ denotes the nonlinear transformation function learned by the NN through the training data. We term this NN-based compensation method as Deep Discriminant Analysis (DDA), for comparison with LDA or NDA.

\section{Experiments}
\subsection{Dataset}\label{sec:datades}
We evaluate the performance of our proposed methods on a short-duration dataset generated from the NIST SRE corpus. This short duration text-independent task is more difficult for speaker verification. The training set consists of selected data from SRE$04$-$08$, Switchboard II phase $2$, $3$ and Switchboard Cellular Part$1$, Part$2$. After removing silence frames using an energy-based VAD, the utterances are chopped into short segments (ranging from $3$-$5$s). The final training set contains $4000$ speakers and each speaker has $40$ short utterances. The enrollment set and test set are derived from NIST SRE $2010$ following a similar procedure. The enrollment set contains $300$ speakers ($150$ males and $150$ females) and each speaker is enrolled by $5$ utterances.  The test set contains $4500$ utterances from the $300$ models in the enrollment set. The trial list we create contains $392660$ trials. There are $15$ positive samples and $1294$ negative samples on average for each model. No cross-gender trial exists.

\subsection{System Details}

\subsubsection{Baseline Settings}
The baseline {\it i}-vector system is implemented using the Kaldi toolkit, 20-dimensional MFCC coefficients with their first and second order derivatives are extracted from the speech segments (identified with an energy based VAD). A 25 ms Hamming window with a 10 ms frame shift is adopted in the feature extraction process. The universal background model (UBM) contains 2048 Gaussian mixtures and the {\it i}-vector dimension is set to 600. Different scoring methods are applied to the length normalized {\it i}-vectors.  3 different scoring methods are adopted to evaluate the performance. Cos denotes the cosine similarity of two vectors, while Euc denotes the Euclidean Distance. As shown in Table \ref{tab:results}, PLDA achieves best performance for the raw input {\it i}-vectors with a EER of 4.96\%, since PLDA itself is both a compensation and scoring method. LDA's dimension in Table \ref{tab:results} is set as 300 and obtains significant performance improvement when applied to Cos or Euc scoring methods. However, no improvement is observed when combining LDA and PLDA.

\subsubsection{Neural Network Settings}

As shown in Table \ref{tab:config}, we adopt a standard feed-forward neural network as the compensation model, which contains one input layer, one hidden layer and one embedding layer. PReLU\cite{he2015delving} is chosen as the activation function, while a batch normalization layer is added before the embedding layer to stabilize the training procedure. The whole network is trained under the joint supervision of softmax loss and center loss, with the value of $\lambda$ in Equation \ref{eq:loss} set as 0.01 (Detailed explanation of this setting can be found in Section \ref{sec:lambda}). Following the strategy used in \cite{wen2016discriminative}, besides the $\lambda$ to balance the impact of two losses, a different learning rate is used for the center loss parameters. The learning rate for the basic neural network is set to 0.01 and the one for center loss is set to 0.1. In the training stage, since it's impractical to update the centers with respect to the whole training set, we update the centers per mini-batch instead, centers are computed by averaging the embeddings of corresponding classes (centers of some classes may not be updated).
\begin{table}[!htb] \small
\caption{Neural Network Configuration}
  \centering
  \begin{tabular}{ c||c|c }
  \hline
    \textbf{Input} & \multicolumn{2}{c}{Source {\it i}-vectors of 600 dimension } \\ \hline \hline
    \textbf{Linear Layers} & \textbf{number of nodes} & \textbf{nonlinear}  \\ \hline
    \textbf{Input Layer} & 600 & PReLU \\
    \textbf{Hidden Layer} & 600 & PReLU + BatchNorm\\ 
    \textbf{Embedding Layer} & 300 & None \\ \hline \hline
    \textbf{Loss} & \multicolumn{1}{c|}{softmax loss} & \multicolumn{1}{c}{0.01 * center loss} \\ \hline
  \end{tabular}
\label{tab:config}
\end{table}

\subsection{Results and Analysis}
\begin{figure}[!ht]
	\centering
	\includegraphics[width=0.9\linewidth]{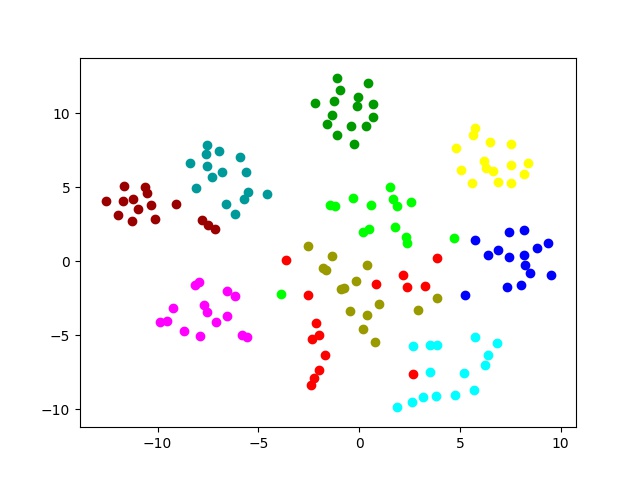}
	\caption{Visualization of {\it i}-vectors}
    \label{fig:ivec}
\end{figure}

\begin{figure}[!ht]
	\centering
	\includegraphics[width=0.9\linewidth]{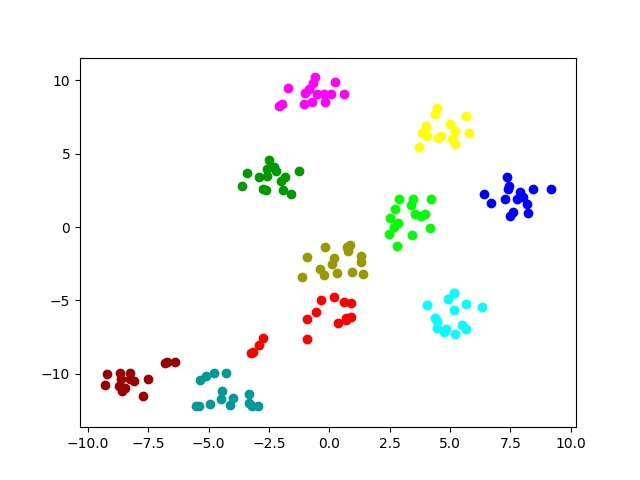}
	\caption{Visualization of compensated embeddings }
    \label{fig:embedding}
\end{figure}

The proposed neural network based system is evaluated on the dataset described in Section \ref{sec:datades}. As shown in Table \ref{tab:results}, compared to LDA, the NN-based DDA obtains larger improvement for Cos and Euc scoring methods, while the best performance of EER 4.69\% is achieved by DDA+Euc, which also outperforms the baseline PLDA system. However, the proposed compensation method is not compatible with PLDA. To better understand the proposed method's effect, we use t-SNE\cite{maaten2008visualizing} to visualize the {\it i}-vectors and their corresponding DDA-compensated embeddings in Fig.\ref{fig:ivec} and Fig.\ref{fig:embedding} (\textbf{Best viewed in color}).
\begin{table}[!ht]
\caption{EER (\%) of different compensation methods}
	\centering
	\begin{tabular}{c c c c}
		\cline{1-4}
		Methods & Cos & Euc & PLDA\\\hline \hline
		Baseline & 7.29 & 6.04 & 4.96 \\ \hline \hline
        LDA & 5.89 & 5.22 & 5.0 \\
	    DDA & 4.78 & \textbf{4.69} & 7.32\\\hline
	\end{tabular}
	\label{tab:results}
\end{table}

Fig.\ref{fig:ivec} depicts the distribution of {\it i}-vectors from 10 speakers randomly chosen from the test set, while the distribution of corresponding compensated embeddings are shown in Fig.\ref{fig:embedding}. As shown in the two figures, with the proposed compensation method, the distribution of embeddings from the same speaker seems more compact, which means the intra-speaker variation is significantly reduced. 

\subsubsection{Impact of the loss weight}\label{sec:lambda}
\begin{figure}[!ht]
	\centering
	\includegraphics[width=0.9\linewidth]{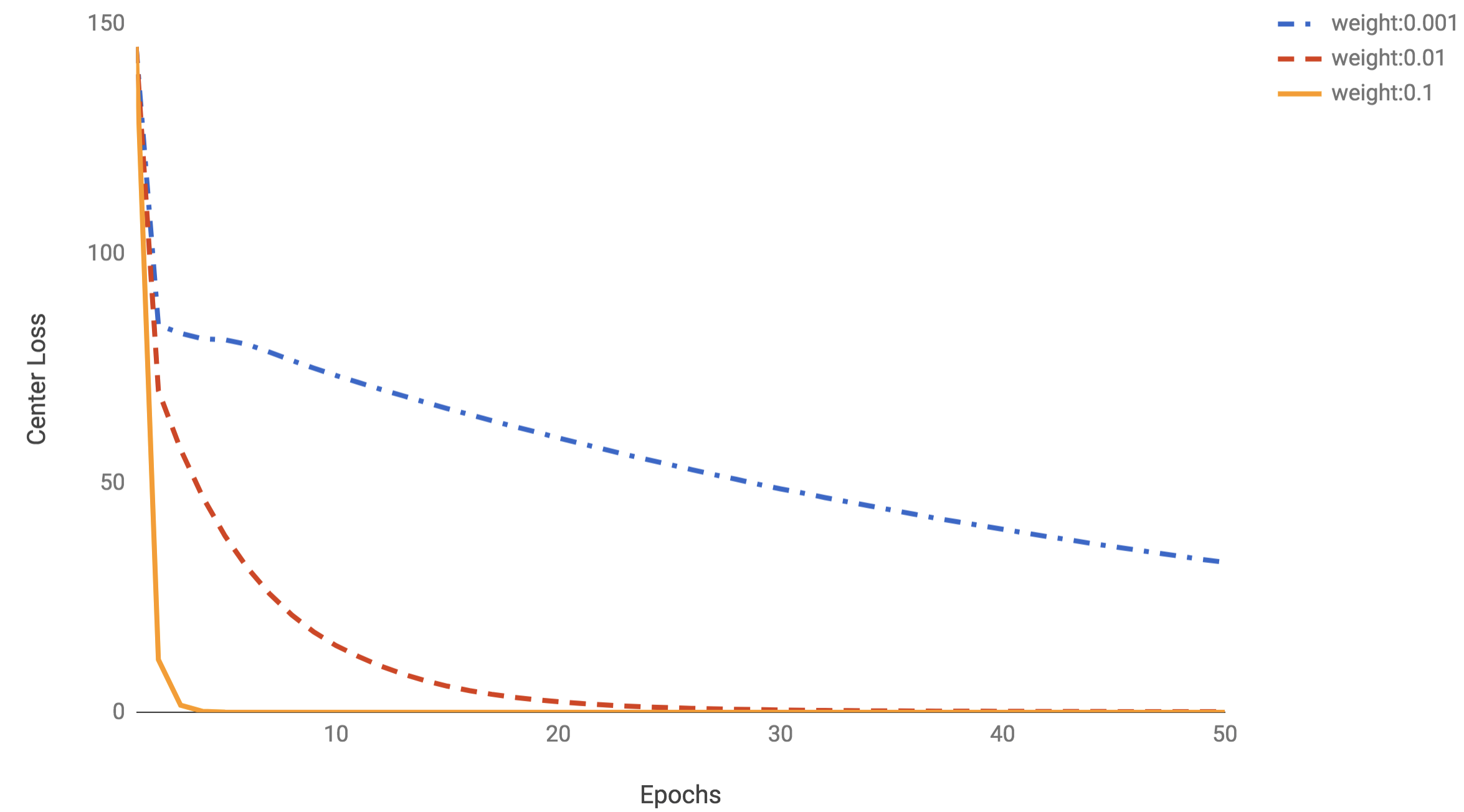}
	\caption{Center Loss with the training epochs}
    \label{fig:centerloss}
\end{figure}

\begin{figure}[!ht]
	\centering
	\includegraphics[width=0.9\linewidth]{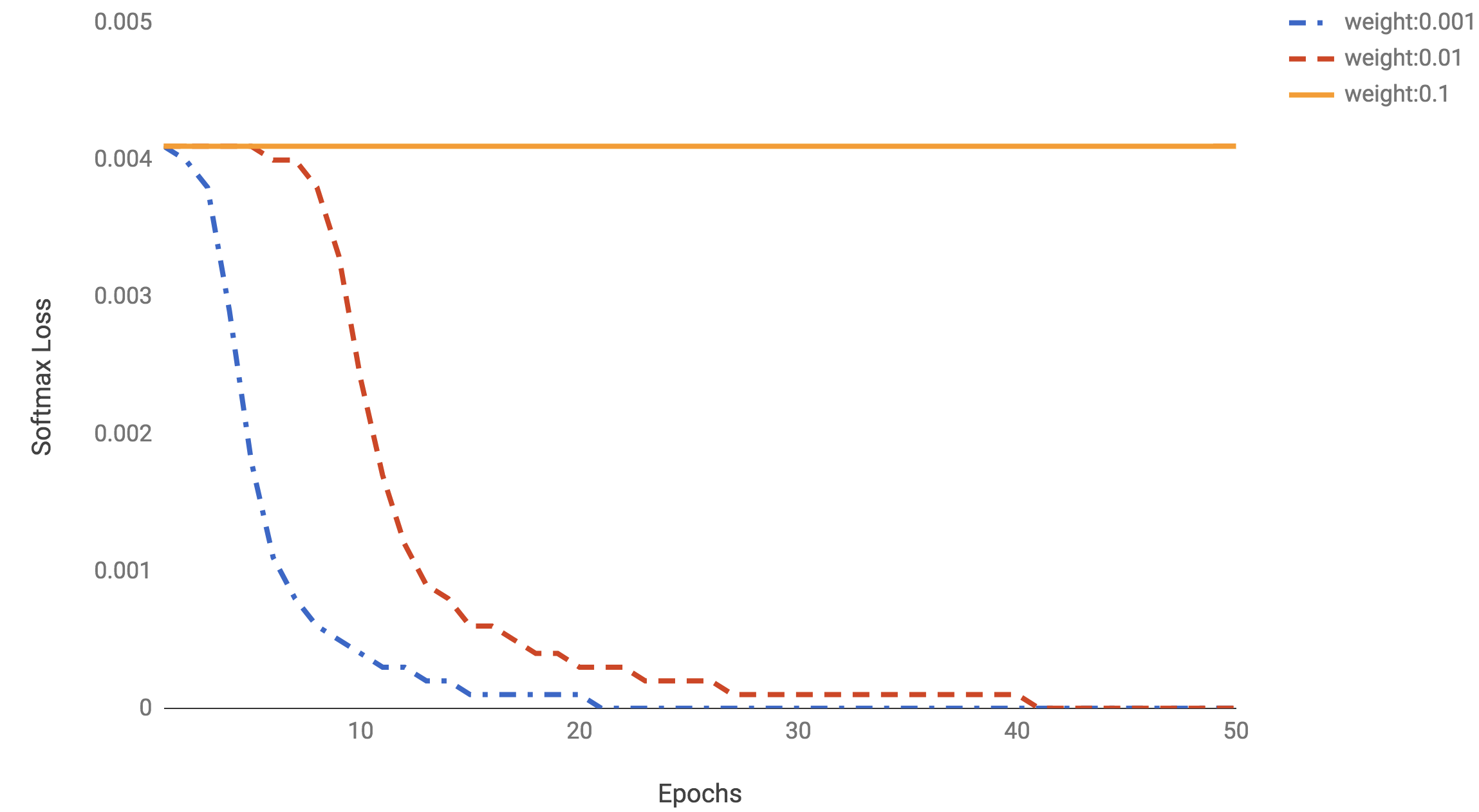}
	\caption{Softmax Loss with the training epochs}
    \label{fig:softmaxloss}
\end{figure}

As mentioned in above sections, a weight $\lambda$ is used to balance the softmax loss and center loss. A small $\lambda$ implies a strong supervision signal provided by the softmax loss, whereas a large $\lambda$ implies a strong supervision signal from the center loss. As shown in Fig.\ref{fig:centerloss} and Fig.\ref{fig:softmaxloss}, when the weight is set to 0.1, the network is actually not trainable, though the center loss degrades quickly, the softmax loss hardly change. In this case, the embeddings are trained to be similar to each other and became not distinguishable. As the value of $\lambda$ is reduced, the softmax loss degrades faster due to its relatively stronger supervision signal. In our experiments, when $\lambda$ varies from 0.001 to 0.01, the training converges faster, while the compensation performance hardly changes.

\subsubsection{Impact of the embedding dimension}
In this section, we investigated the impact of different dimensions of the projection subspace by varying the embedding layer's dimension. As shown in Table \ref{tab:results_dim}, it's interesting to find that DDA achieves best performance with 400 dimension, in contrary, LDA achieves best performance with 200 dimension. Though not listed in the following table, it should be mentioned that with the dimension of 100 or 500 which are not listed, the EER increases for both two compensation methods.

\begin{table}[!ht] 
\caption{EER (\%) comparison }
	\centering
	\begin{tabular}{c | c| c c c}
		\cline{1-5}
        Scoring & Compensation & 200dim & 300dim & 400dim \\\hline \hline
        \multirow{2}{*}{Cos} & LDA & 5.53 & 5.89 & 6.28 \\
	    & DDA & 5.31 & 4.78 & 4.67\\\hline
        \multirow{2}{*}{Euc} & LDA & 5.22 & 5.22 & 5.40 \\
	    & DDA & 5.08 & 4.69 & \textbf{4.51}\\\hline
	\end{tabular}
	\label{tab:results_dim}
\end{table}

\section{Conclusion}
Intra-speaker variability compensation techniques such as LDA have been researched a lot in the state-of-the-art {\it i}-vector framework, LDA has several limitations dual to the mismatch between LDA's assumptions and the true distribution of {\it i}-vectors. In this paper, we proposed a non-linear compensation framework based on a discriminative neural network, termed as DDA (Deep Discriminant Analysis). The neural network is trained under the joint supervision of softmax loss and center loss, the softmax loss forces the learned embeddings of different classes staying apart, while the center loss pulls the embeddings from the same class close to their centers. Experiments shows that with the assistance of the proposed compensation method, simple Cosine Scoring or Euclidean Scoring can achieve even better performance than PLDA. 



\bibliographystyle{IEEEbib}
\bibliography{Odyssey2018_BibEntries}

\end{document}